\documentclass[12pt]{iopart}
\usepackage{iopams}  
\usepackage{bm}
\usepackage{graphicx}
\begin{document}

\title[Topological electronic states and thermoelectric transport in single-layer WSe$_2$]{
Topological electronic states and thermoelectric transport at phase boundaries in single-layer WSe$_2$: An effective Hamiltonian theory
}

\author{G. Tkachov}

\address{Institute of Physics, Augsburg University, 86135 Augsburg, Germany}
\ead{gregor.tkachov@physik.uni-augsburg.de}
\vspace{10pt}
\begin{indented}
\item[]August 2020
\end{indented}

\begin{abstract}
Monolayer transition metal dichalcogenides in the distorted octahedral 1T$^\prime$ phase 
exhibit a large bulk bandgap and gapless boundary states, which is an asset in 
the ongoing quest for topological electronics. 
In single-layer tungsten diselenide (WSe$_2$), the boundary states have been observed 
at well ordered interfaces between 1T$^\prime$ and semiconducting (1H) phases. 
This paper proposes an effective 4-band theory for the boundary states in single-layer WSe$_2$,
describing a Kramers pair of in-gap states as well as the behaviour at 
the spectrum termination points on the conduction and valence bands of the 1T$^\prime$ phase.
The spectrum termination points determine the temperature and chemical potential dependences 
of the ballistic conductance and thermopower at the phase boundary.
Notably, the thermopower shows an ambipolar behaviour, changing the sign in the bandgap of the 1T$^\prime$ - WSe$_2$ 
and reflecting its particle-hole asymmetry. 
The theory establishes a link between the bulk band structure and ballistic boundary transport in single-layer WSe$_2$ 
and is applicable to a range of related topological materials.
\end{abstract}

%
%
%
%
%

\section{introduction}

Topological defects interpolating between distinct quantum ground states can host localized fermions, 
harboring rich physics (see, e.g., Refs. \cite{Jackiw76,Su79,Kitaev01,Nayak08}).
In solids, nontrivial topology can emerge from an inversion of electronic bands 
at the interface of two materials \cite{Volkov85} or from an inverted band ordering in ${\bm k}$ space \cite{Kane05,Bernevig06,Murakami06}, 
leading in each case to localized boundary states.
In particular, two-dimensional topological insulators (2DTIs) with strong spin-orbit-coupling (SOC) \cite{Kane05,Bernevig06,Murakami06,Koenig07,Knez11} 
possess a pair of edge states related by time reversal and existing in the bandgap of the material.

Recently, monolayer transition metal dichalcogenides in the distorted octahedral 1T$^\prime$ phase 
have been predicted to be 2DTIs with an intrinsic inverted band structure \cite{Qian14}.
Experimentally, the edge states have been reported in single-layer tungsten ditelluride (WTe$_2$) \cite{Fei17,Wu18} 
and tungsten diselenide (WSe$_2$) \cite{Ugeda18}. In the latter case, the topological edge states come as boundary states  
at the crystallographically aligned interface between a 1T$^\prime$ phase domain and a semiconducting 1H domain of WSe$_2$.
Crystalline phase boundaries in WSe$_2$ are well ordered, accessible to high-resolution scanned probe microscopy and    
offer other opportunities for testing predictions regarding topological edge states (see recent review in \cite{Culcer20}). 
In the ongoing quest for topological electronics, transport properties of WSe$_2$ phase interfaces deserve particular attention.   

There is also a general theoretical reason for taking a closer look at the boundary states in WSe$_2$. 
In typical 2DTIs \cite{Kane05,Bernevig06}, the edge states resemble massless Dirac fermions in the sense 
that they exhibit a linear level crossing over a substantial energy range. 
In this energy range, the essential properties of the edge states, such as their electric transport, 
can be successfully explained by a Dirac-like model.  
In contrast, in WSe$_2$, the crossing of the boundary states is highly nonlinear \cite{Ugeda18}, 
rendering the picture of the Dirac fermions invalid in this case.  
Consequently, transport calculations based on Dirac-like models cannot be directly applied to the boundary states in WSe$_2$. 

This paper examines the boundary states in 2D WSe$_2$, using an effective Hamiltonian theory for 1T$^\prime$ - 1H phase boundaries.
The effective Hamiltonian operates in a reduced Hilbert space spanned by the conduction and valence bands, including the spin, as adopted in \cite{Shi19}. 
We find a strongly nonlinear boundary spectrum reminiscent of a SO-split parabolic band in a 1D conductor. 
Its nonlinearity and particle-hole asymmetry are consistent with the ab initio calculations of Ugeda {\em et al} \cite{Ugeda18}. 

The solution for the boundary states is implemented to calculate their electric conductance and thermopower in the ballistic regime.
A subtlety is that the ballistic transport depends on how the boundary spectrum merges into the bulk bands.
This happens at special points on the bulk conduction and valence bands at which the bound state ceases to exist.
The implications of such spectrum termination points for electron transport have not been fully understood yet. 
This question is clarified here for 1T$^\prime$ - WSe$_2$ in the context of the recent experimental and ab initio study \cite{Ugeda18}. 
Notably, the temperature and chemical potential dependences of both conductance and thermopower 
are found to be sensitive to the termination points of the boundary spectrum.
Furthermore, through the spectrum termination points the thermoelectric coefficients depend on 
a structural inversion asymmetry, providing extra information on the material properties.
These results establish a link between the bulk band structure of 1T$^\prime$  - WSe$_2$ and the boundary electron transport, 
and complement the earlier transport studies of 2DTI systems (see, e.g., Refs. \cite{Takahashi10,Ghaemi10,Takahashi12,Xu14,Xu17,Gusev19}). 
The following sections explain the details of the calculations and provide an extended discussion of the results.

\section{Effective Hamiltonian description of mixed-phase 2D ${\rm WSe}_2$}

\subsection{1T$^\prime$ phase. Intrinsic band inversion}
\label{Model}

To set the scene, we define the Hamiltonian for a plane-wave state with wave vector ${\bm k}=[k_x, k_y, 0]$ in a homogeneous 1T$^\prime$ phase,

\begin{equation}
H({\bm k}) = H_0({\bm k}) + H_{\rm SO}({\bm k}).
\label{H_k}
\end{equation}
Here, the first term describes a monolayer without spin-orbit coupling (SOC), 
while the second term accounts for SOC due to a structural $z \to -z$ reflection asymmetry.
As long as only the properties of the conduction and valence bands are of concern, we can work in the reduced Hilbert space in which a 
state vector $|{\bm k}\rangle$ has four components, 
$|{\bm k}\rangle = [ \Psi_{\rm c \uparrow }({\bm k}), \Psi_{\rm v \uparrow}({\bm k}), \Psi_{\rm c \downarrow}({\bm k}), \Psi_{\rm v \downarrow}({\bm k})]^T$,  
where subscripts ${\rm c}$ and ${\rm v}$ refer to the conduction and valence bands, while $\uparrow$ and $\downarrow$ to the spin states.
A Hamiltonian acting in this reduced space -- an effective Hamiltonian -- can be represented by a $4\times 4$ matrix \cite{Qian14,Shi19,Kormanyos15}. 
In particular (see \cite{Shi19}),

\begin{eqnarray}
H_0({\bm k}) &=& 
\left[
\begin{array}{cccc}
 \epsilon_{\bm k} + m_{\bm k}  &  v_x k_x + iv_y k_y & 0 & 0 \\
 v_x k_x - iv_y k_y &   \epsilon_{\bm k} - m_{\bm k} & 0 & 0 \\
 0 & 0 &  \epsilon_{\bm k} + m_{\bm k} & - v_x k_x + iv_y k_y  \\
 0 & 0  & - v_x k_x - iv_y k_y & \epsilon_{\bm k} - m_{\bm k}
 \end{array}
\right] 
\nonumber\\
&=&
\epsilon_{\bm k}\sigma_0\tau_0  + v_x k_x \sigma_z\tau_1 - v_y k_y \sigma_0 \tau_2 + m_{\bm k} \sigma_0 \tau_3,
\label{H_0}
\end{eqnarray}
where $\epsilon_{\bm k}$ and $m_{\bm k}$ are quadratic functions of the wave vector given by
\begin{equation}
\epsilon_{\bm k}= \epsilon_0 +  \epsilon_x k^2_x +\epsilon_y k^2_y, \qquad m_{\bm k}= m_0 +  m_x k^2_x +m_y k^2_y, 
\label{m_k}
\end{equation}
and $\epsilon_0$, $m_0$, $\epsilon_{x,y}$, $m_{x,y}$, and $v_{x,y}$ are the band structure constants of the effective model.
In particular, $\epsilon_0 \pm m_0$ are the energies of the conduction and valence bands at the $\Gamma$ point; 
$\epsilon_{x,y}$  and $m_{x,y}$ characterize the band curvature, while $v_{x,y}$ the atomic SOC.
We use the Pauli matrices in the band ($\tau_1, \tau_2$, and $\tau_3$) and spin ($\sigma_z$) subspaces 
along with the corresponding unit matrices $\tau_0$ and $\sigma_0$. 

The lack of the $z \to - z$ symmetry allows for $H_{\rm SO}({\bm k})$ of different types. 
We consider a particular one

\begin{equation}
H_{\rm SO}({\bm k}) = 
\left[
\begin{array}{cccc}
 \lambda k_y &   i\delta & 0 & 0 \\
- i\delta &   \lambda k_y & 0 & 0 \\
 0 & 0 &  -\lambda k_y & -i\delta  \\
 0 & 0  &  i\delta & -\lambda k_y
 \end{array}
\right]
= (\lambda k_y \tau_0 - \delta \tau_2)\sigma_z,
\label{H_SO}
\end{equation}
where $\lambda$ and $\delta$ are the structural SOC constants. 
The term specified by equation (\ref{H_SO}) can originate from an applied out-of-plane electric field
which controls the coupling constants $\lambda$ and $\delta$ \cite{Shi19}. For the purpose of this study, 
it is sufficient that the block-diagonal $H_{\rm SO}({\bm k})$ (\ref{H_SO}) lifts the spin degeneracy of $H_0({\bm k})$ (\ref{H_0}).
The inclusion of the off-diagonal SOC would result in a more involving boundary problem later on. 

\begin{figure}[t]
\begin{center}
\includegraphics[width=85mm]{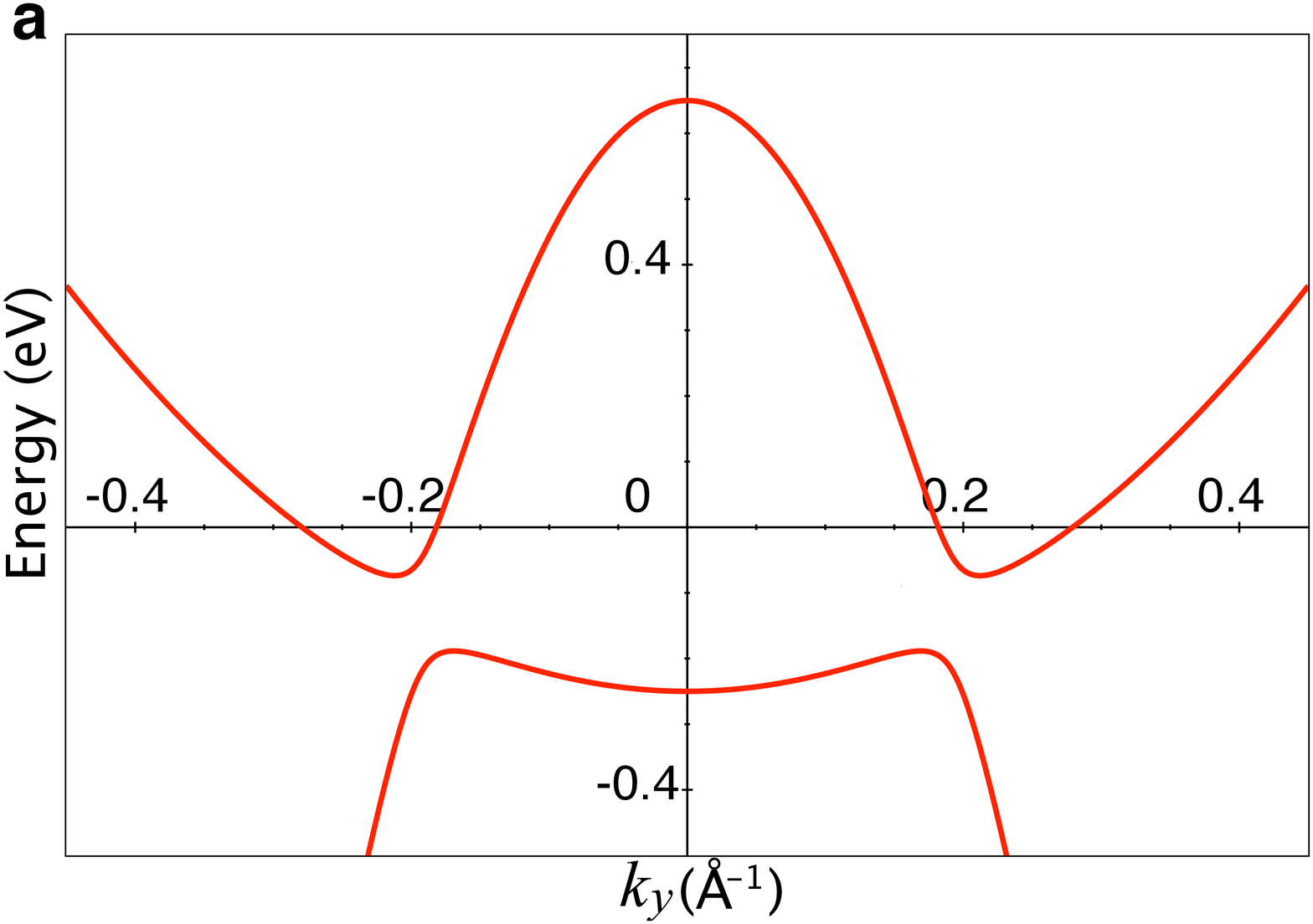}
\includegraphics[width=70mm]{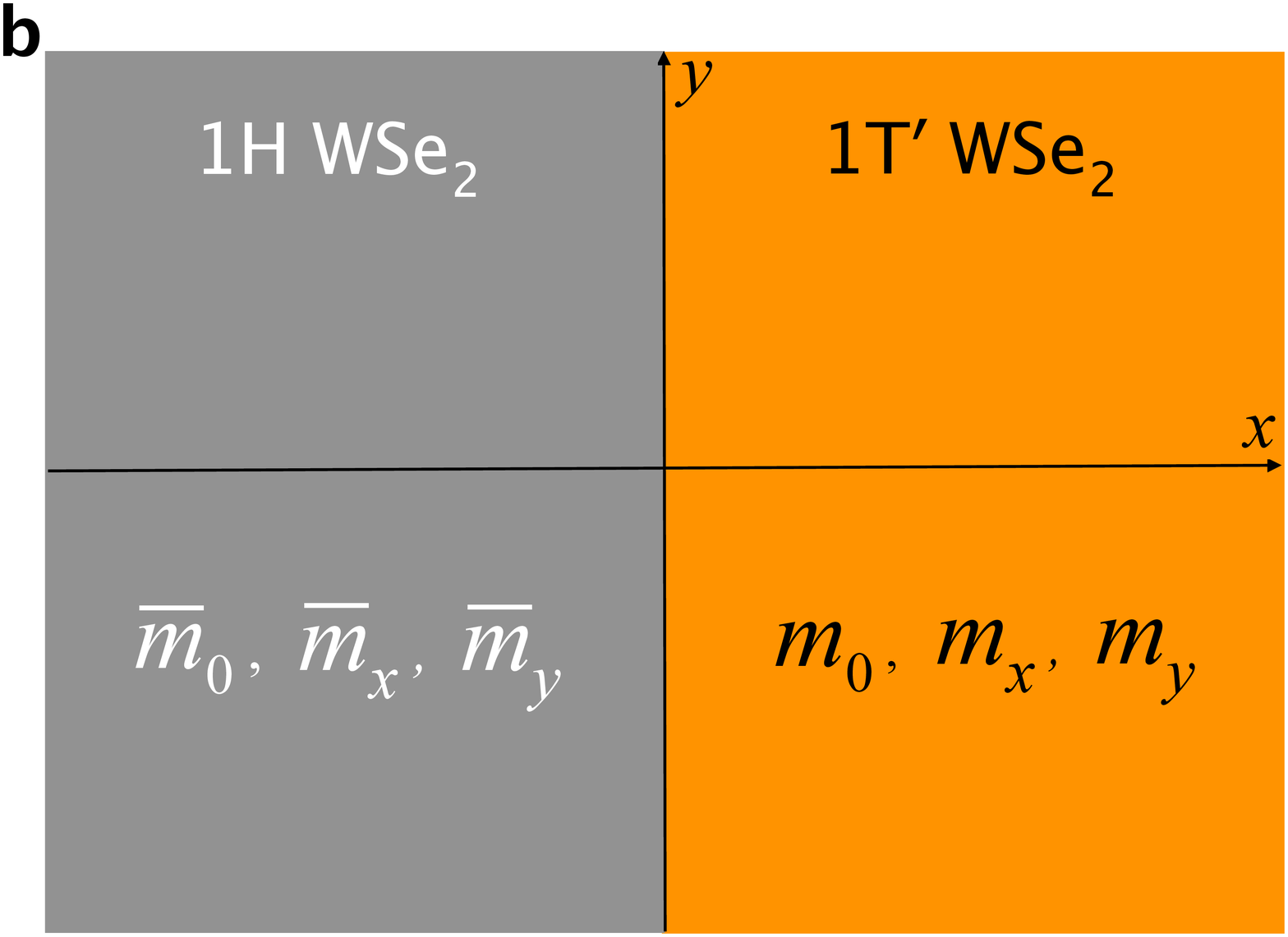}
\caption{\label{E_WSe2}
(a)
Conduction and valence bands of single-layer 1T$^\prime$ - WSe$_2$ from the effective Hamiltonian model.  
The plot shows the band dispersion in equation (\ref{E_bulk}) along the $k_y$ direction for 
$\epsilon_0 = 0.2$ eV, $\epsilon_x= -10$ eV${\rm \AA}^2$, $\epsilon_y= -9$ eV${\rm \AA}^2$, $v_x = 0.8$ eV${\rm \AA}$, $v_y = 0.45$ eV${\rm \AA}$, 
$m_0 = 0.45$ eV, $m_x = -13$ eV${\rm \AA}^2$, $m_y = -12$ eV${\rm \AA}^2$, and $\lambda =\delta=0$.
The effective model is tailored to qualitatively reproduce the ab initio band structure calculations along the $Y-\Gamma-Y$ direction in the Brillouin zone of 
1T$^\prime$ - WSe$_2$ \cite{Ugeda18}, e.g. the bandgap $E_g \approx 120$ meV.
 Along the $k_x$ direction, the band dispersion is similar, with a somewhat larger bandgap.
(b)
Schematic of a 1T$^\prime$ - 1H phase boundary in the effective model of WSe$_2$ (see also equation (\ref{m(x)}) and Table \ref{Table}).
}
\end{center}
\end{figure}

The band structure of the effective model is given by the eigenvalues of the total Hamiltonian (\ref{H_k}). 
It is instructive to look at the dispersion of the conduction and valence bands:

\begin{eqnarray}
E^\pm_\sigma({\bm k}) &=& \epsilon_0 + \epsilon_x k^2_x + \epsilon_y k^2_y + \lambda \sigma k_y 
\nonumber\\
& \pm &
\sqrt{
 v^2_xk^2_x + ( v_yk_y + \delta\sigma )^2 + (m_0 + m_x k^2_x + m_y k^2_y)^2,
}
\label{E_bulk}
\end{eqnarray}
where $\sigma = \pm 1$ is the eigenvalue of $\sigma_z$. 
With appropriately chosen parameters, equation (\ref{E_bulk}) qualitatively reproduces 
the conduction and valence bands of single-layer 1T$^\prime$ - WSe$_2$ (see figure \ref{E_WSe2}(a)).
The positions of the bands, their profiles and the energy gap between them 
overall agree with the ab initio calculations along the $Y-\Gamma-Y$ direction in the Brillouin zone (cf. \cite{Ugeda18}). 
The effective model is tailored to have the bandgap  $E_g \approx 120$ meV as calculated in  \cite{Ugeda18}.
Notably, the band curvature at $k_y=0$ ($\Gamma$ point) indicates an inverted band ordering (cf. \cite{Qian14}), 
which is a necessary prerequisite for the occurrence of the topological boundary modes. 
As in the Bernevig-Hughes-Zhang (BHZ) model \cite{Bernevig06},
an intrinsic band inversion is realized under conditions $m_0m_x < 0$ and $m_0m_y < 0$ for the coefficients 
in the gap term $m_{\bm k}$ in equation (\ref{m_k}), see also the caption for figure \ref{E_WSe2}(a).

\subsection{ 1T$^\prime$ - 1H phase interface. Topological boundary states}
\label{Boundary states}

The experiment of Ugeda {\em et al} \cite{Ugeda18} dealt with a crystallographically well-defined interface between a 1T$^\prime$ phase domain  
and a semiconducting 1H domain in contiguous single layers of WSe$_2$. 
We assume that the two phases are separated by a straight boundary,
choosing the $x$ and $y$ axes perpendicular and parallel to it, as shown in figure \ref{E_WSe2}(b). 
In this geometry, $k_y$ remains a good quantum number, while $k_x$ needs to be replaced by the operator $-i\partial_x$.
Compared to the 1T$^\prime$ phase, the 1H one has a larger bandgap and a normal band ordering. 
This difference can be accounted for by an appropriately generalized gap term $m_{\bm k}$. 
To model the 1T$^\prime$ - 1H interface, we use the position-dependent gap term

\begin{equation}
m_{\bm k} \to m_0(x) - m_x(x)\partial^2_x+ m_y(x) k^2_y, 
\label{m(x)}
\end{equation}
where the coefficients $m_0(x)$ and $m_{x,y}(x)$ coincide with $m_0$ and $m_{x,y}$ in the 1T$^\prime$ domain ($x\geq 0$),
while taking different values $\overline{m}_0$ and $\overline{m}_{x,y}$ on the 1H side ($x\leq 0$). 
The relative sign of $\overline{m}_0$ and $\overline{m}_{x,y}$ is positive, meaning a normal band ordering in the 1H domain
(see also Table \ref{Table} summarizing the effective interface model).

\begin{table}
\caption{Parametrization of the gap term (\ref{m(x)}) in the model of the 1T$^\prime$ - 1H WSe$_2$ interface.
The relative sign of $m_0$ and $m_{x,y}$ (resp. $\overline{m}_0$ and $\overline{m}_{x,y}$) corresponds to an inverted (resp. normal) band ordering in the 1T$^\prime$ (resp. 1H) phase.}
\begin{indented}
\item[]
\begin{tabular}{@{}ccccc}
\br
                                                       & $m_0(x)$  & $m_x(x)$   & $m_y(x)$    &     Band structure ordering     \\ 
\mr
1T$^\prime$ domain ($x\geq 0$)   & $m_0$    & $m_x$     & $m_y$      & inverted ($m_0m_{x,y} < 0$)    \\

1H domain ($x\leq 0$)                   & $\overline{m}_0$  & $\overline{m}_x$   & $\overline{m}_y $    & normal ($\overline{m}_0\overline{m}_{x,y} > 0$)   \\
\br
\end{tabular}
\end{indented}
\label{Table}
\end{table}

Using equations (\ref{H_k}) -- (\ref{H_SO}) and (\ref{m(x)}), we can write the interface Hamiltonian as

\begin{equation}
H_{\sigma k_y}(x) = H^{^A}_{\sigma k_y}(x) + H^{^S}_{\sigma k_y}(x),
\label{H_x}
\end{equation}
where $H^{^S}_{\sigma k_y}(x)$ and $H^{^A}_{\sigma k_y}(x)$ are the particle-hole symmetric and asymmetric parts of the Hamiltonian. 
For given spin direction $\sigma = \pm 1$ (resp. $\uparrow$ and $\downarrow$), 
the two terms in equation (\ref{H_x}) are $2\times 2$ matrices given by
 
\begin{equation}
H^{^A}_{\sigma k_y}(x) = (\epsilon_0 + \epsilon_y k^2_y + \lambda \sigma k_y  - \epsilon_x \partial^2_x) \tau_0
\label{H_A}
\end{equation}
and

\begin{equation}
H^{^S}_{\sigma k_y}(x) = -i\sigma v_x \partial_x \tau_1 - (v_y k_y + \delta\sigma) \tau_2 + [m_0(x) + m_y(x) k^2_y - m_x(x)\partial^2_x] \tau_3.
\label{H_S}
\end{equation}
This yields the eigenvalue equation

\begin{equation}
H^{^A}_{\sigma k_y}(x)|x\rangle + H^{^S}_{\sigma k_y}(x)|x\rangle =E |x\rangle
\label{Eq_E}
\end{equation}
for energy $E$ and a real-space two-component wave function $|x\rangle$. 
The latter is assumed to vanish away from the interface: $|x\rangle \to 0$ for $x \to \pm\infty$, 
while being continuous at $x=0$.  

The above boundary problem is solved in \ref{Boundary solution}. The result for the wave function is

\begin{equation}
|x\rangle = |0\rangle \cases{
-\frac{\varkappa_3 + \varkappa_2}{\varkappa_1 - \varkappa_2} {\rm e}^{-\varkappa_1 x} 
+ \frac{\varkappa_3 + \varkappa_1}{\varkappa_1 - \varkappa_2}{\rm e}^{-\varkappa_2 x},  & for $x \geq 0$,\\
 {\rm e}^{\varkappa_3 x}, & for $x \leq 0$.\\}
\label{Sol_x}
\end{equation}
It describes a bound state localized on the length-scales $\varkappa^{-1}_{1,2}$ and $\varkappa^{-1}_3$
in the 1T$^\prime$ and 1H domains, respectively, where

\begin{equation}
\varkappa_{1,2} = \left| \frac{ v_x }{2m_x }\right| \pm \sqrt{ \left( \frac{ v_x }{2m_x } \right)^2 + \frac{ m_0 + m_y k^2_y}{ m_x } }
\label{kappa_12_top}
\end{equation}
and

\begin{equation}
\varkappa_3 = {\rm sgn}(m_0 \overline{m}_0) \left|\frac{ v_x }{2\overline{m}_x }\right| + 
\sqrt{ \left( \frac{ v_x }{2\overline{m}_x } \right)^2 + \frac{ \overline{m}_0 + \overline{m}_y k^2_y}{ \overline{m}_x } }.
\label{kappa_3}
\end{equation}
The inverted 1T$^\prime$ band structure allows the $k_y$ values in the segment

\begin{equation}
-k_0 \leq k_y \leq k_0, \qquad k_0 = \sqrt{-m_0/m_y},
\label{k_0}
\end{equation}
where the endpoints $\pm k_0$ are the zeros of the gap term $m_0 + m_y k^2_y$ in equation (\ref{kappa_12_top}). 
Further, it can be shown that the wave function at the boundary, $|0\rangle$, is an eigenstate of the Pauli matrix $\tau_2$  
defined by  

\begin{equation}
\tau_2 |0\rangle = -\sigma \, {\rm sgn}(v_x m_x)  |0\rangle,
\end{equation}
i.e. the choice of the eigenstate depends on the spin projection as well as on the relative sign of the band structure parameters $v_x$ and $m_x$. 
For $\sigma =\pm 1$, there are two orthogonal boundary modes. Their energy dispersion is given by

\begin{equation}
E_{\sigma k_y} = \epsilon + \epsilon_y k^2_y + v \sigma k_y,  
\label{Sol_E}
\end{equation}
with

\begin{equation}
v = v_y \, {\rm sgn}(v_x m_x) + \lambda, \qquad \epsilon = \epsilon_0 + \delta \, {\rm sgn}(v_x m_x).
\label{v_epsilon}
\end{equation}
The parameters $v$ and $\epsilon$ absorb the structural SOC constants and account for the signs of other involved parameters 
(see \ref{Boundary solution}).

Overall, the above boundary solution is analogous to the edge states of a 2DTI in the BHZ model.
There are a few new details, though. The solution obtained for the BHZ model (see, e.g., \cite{Zhou08}) is a "hard-wall" one, 
i.e. the electronic wave function vanishes upon approaching the boundary of a 2DTI. 
In contrast, equation (\ref{Sol_x}) accounts for the leakage of the wave function into the semiconducting (1H) 
region, which was observed in the scanning tunneling experiment of Ugeda {\em et al} \cite{Ugeda18}. 
Further, the SOC (\ref{H_SO}) makes the bulk bands asymmetric with respect to $k_y \to -k_y$.  
In this case, we find a specific dependence of the boundary spectrum on the SOC constants $\lambda$ and $\delta$. 
The implications of this finding for the thermoelectric coefficients will be discussed in the next section. 
Finally, the dependence on the signs of the model parameters is generic, 
allowing for different types of band structures.

\begin{figure}[t]
\begin{center}
\includegraphics[width=110mm]{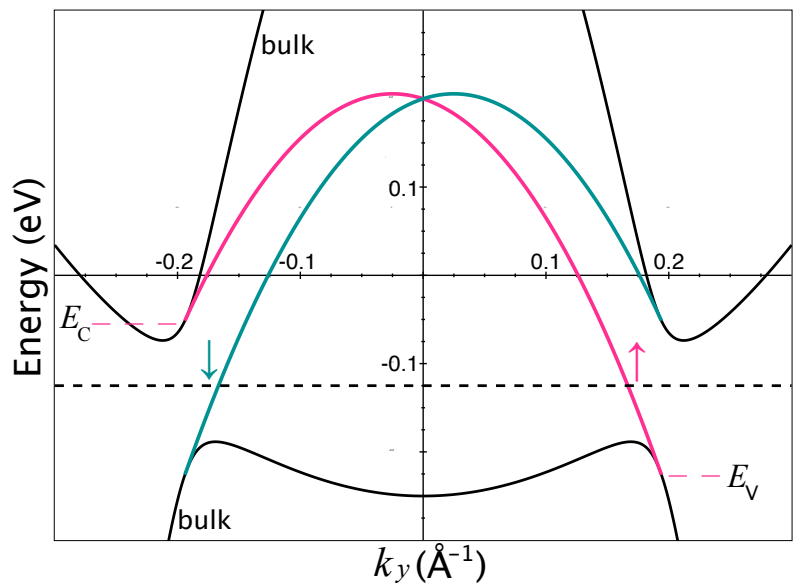}
\end{center}
\caption{
Energy dispersion of boundary modes with spin projections $\uparrow$ and $\downarrow$, see equation (\ref{Sol_E}), 
for the same parameters as in figure \ref{E_WSe2}(a). The boundary modes disperse between the bulk conduction and valence bands
of 1T$^\prime$ - WSe$_2$, forming a Kramers pair in the bandgap.
$E_{\rm c}$ and $E_{\rm v}$ are the energies at the termination points ($\approx \pm 0.2$ ${\rm \AA}^{-1}$) of the boundary spectrum. 
The dashed line indicates the Fermi level adjusted in the bandgap of 1T$^\prime$ - WSe$_2$.
}
\label{E_Interface}
\end{figure}

Figure \ref{E_Interface} shows the energies of the boundary states (see equation (\ref{Sol_E})) 
along with the bulk bands of 1T$^\prime$ - WSe$_2$.
Two boundary modes with opposite spins $\uparrow$ and $\downarrow$ connect the bulk conduction and valence bands, 
crossing at the $\Gamma$ point. Their dispersion resembles a SOC - split parabolic band in a one-dimensional conductor. 
However, the boundary modes terminate on the conduction and valence bands, so only one Kramers pair occurs in the bandgap.
The termination points of the boundary spectrum are the endpoints of the allowed $k_y$ segment in equation (\ref{k_0}). 
In figure \ref{E_Interface}, $\pm k_0$ are approximately $\pm 0.2$ ${\rm \AA}^{-1}$. 
At these points the bound state (\ref{Sol_x}) gets delocalized, spreading into the bulk of the 1T$^\prime$ domain. 
In energy, the spectrum termination points lie at
\begin{equation}
E_{\rm c, v} = \epsilon + \epsilon_y k^2_0 \mp v k_0 = \epsilon - \frac{\epsilon_y m_0}{m_y} \mp v \sqrt{ - \frac{m_0}{m_y} }
\label{E_cv}
\end{equation}
in the conduction ("$-$") and valence ("$+$") band, respectively. 
In figure \ref{E_Interface}, these energies are $E_{\rm c} \approx -50$ meV and  $E_{\rm v} \approx -225$ meV.
The spectrum termination points $\pm k_0$ do not coincide with the positions of the local extrema of the bulk bands,
so the energy difference $E_{\rm c} - E_{\rm v} \approx 175$ meV is somewhat larger than the bandgap $E_g \approx 120$ meV,
although the scale is the same. 

Beside the energy spectrum, the above results provide an estimate for the distance over which the boundary states decay from the interface. 
In the 1T$^\prime$ domain, the decay length is of order $|2m_x/v_x|$ (see equation (\ref{kappa_12_top})), which for the chosen parameters is 
about $3$ nm. For comparison, the estimate of Ugeda {\em et al} \cite{Ugeda18} is $2$ nm. 

It is also worth mentioning that a matrix element between the Kramers partners $|k_y, \uparrow\rangle$ and $|-k_y, \downarrow\rangle$ 
satisfies the relation

\begin{equation}
\langle -k_y, \downarrow | V |k_y, \uparrow \rangle = - \langle -k_y, \downarrow | V^\dagger |k_y, \uparrow\rangle
\label{Kramers}
\end{equation}
valid for a local operator $V$ commuting with the time-reversal operator $\mathbb{T} = -i\sigma_y K$ ($\dagger$ and $K$ denote hermitian and complex conjugations). As a consequence, a hermitian potential preserving the time-reversal symmetry 
causes no backscattering of the boundary modes in the bandgap 
because the matrix element $\langle -k_y, \downarrow | V |k_y, \uparrow\rangle$ vanishes identically in that case. 
The mean free path of the boundary states can be limited by elastic spin-flip scattering with a potential $V \not = \mathbb{T} V \mathbb{T}^{-1}$. 
In this case, spin-flip scattering is formally analogous to the intervalley scattering of edge states in spinless graphene \cite{GT12} 
and can be treated by the same methods. 
For example, in the self-consistent Born approximation, the intervalley scattering determines the transport mean free path of the edge states, 
while intravalley scattering only contributes to the quasiparticle life-time \cite{GT12}. A similar situation can be expected for a phase boundary 
in WSe$_2$ in the presence of elastic spin-flip scatterers.

\section{Electric conductance and thermopower of the boundary states}

When the Fermi level is adjusted in the bandgap of the 1T$^\prime$ domain (see also figure \ref{E_Interface}), 
the phase boundary acts as a quasi-1D conductor with a Kramers pair of propagating modes.
We discuss first the equilibrium case. For one mode, say the $\uparrow$ one, the electric current can be calculated as 
the equilibrium expectation value in $k$ space: 

\begin{equation}
j_\uparrow = e \int_{-k_0}^{k_0}  \nu_{\uparrow}(k_y) f[E_\uparrow(k_y)] \frac{dk_y}{2\pi}= 
e \int_{E_{\rm v}}^{E_{\rm c}} N_{\uparrow}(E) \nu_{\uparrow}(E) f(E) dE,
\label{j_up}
\end{equation}
where we trace over all $k_y$ values of the boundary mode (see equation (\ref{k_0})), 
$\nu_{\uparrow}(k_y) = \hbar^{-1} \partial E_{\uparrow}(k_y) / \partial  k_y$ is the mode velocity, and 
$f(E) = \bigr[ {\rm e}^{(E-\mu)/(k_{_B}T)} + 1 \bigl]^{-1}$ is the Fermi occupation number 
($\mu$ and $k_{_B}$  are the chemical potential and Boltzmann constant)
\footnote{
A recent lattice-model study of equilibrium boundary currents has been reported in 
Wei Chen, Phys. Rev. B {\bf 101}, 195120 (2020).
}. 
The $k_y$ integration is replaced by the energy integral with the 1D density of states (DOS) $N_{\uparrow}(E) = h^{-1} |\nu_{\uparrow}(E)|^{-1}$ and 
velocity
\begin{equation}
\nu_\uparrow(E) = \frac{ {\rm sgn}(\epsilon_y) }{\hbar} \sqrt{  v^2 - 4\epsilon_y (\epsilon - E) },
\label{nu_up}
\end{equation}
obtained from equation (\ref{Sol_E}). Only the energies between the spectrum termination points $E_{\rm v}$ and $E_{\rm c}$ (\ref{E_cv}) 
contribute to the current because this energy window corresponds to a chiral (one-way moving) state 
\footnote{
In the energy interval from $E_{\rm c}$ to the top of the boundary band (see figure \ref{E_Interface}), 
the boundary spectrum is symmetric with respect to the position of the maximum. 
Therefore, this energy interval does not contribute to the current in equation (\ref{j_up}).
}. 
The sign of its velocity, ${\rm sgn}(\nu_{\uparrow}) =  {\rm sgn}(\epsilon_y)$, 
determines the direction of the current:
  
\begin{equation}
j_\uparrow = \frac{e}{h}  \int_{E_{\rm v}}^{E_{\rm c}} {\rm sgn}(\nu_{\uparrow}) f(E) dE = 
{\rm sgn}(\epsilon_y) \frac{e}{h}  \int_{E_{\rm v}}^{E_{\rm c}} f(E) dE.
\label{j_up1}
\end{equation}
At zero temperature, the current $j_\uparrow = {\rm sgn}(\epsilon_y) (e/h) (\mu - E_{\rm v})$ is carried by all occupied states from  $E_{\rm v}$
to $\mu$. Likewise, the current $j_\downarrow$ depends on the sign of the velocity of the $\downarrow$ mode, 
${\rm sgn}(\nu_{\downarrow}) =  -{\rm sgn}(\epsilon_y)$, which is opposite to that in equation (\ref{j_up1}).
At equilibrium, the two modes are equally occupied, rendering the net electric current $j = j_\uparrow + j_\downarrow$ null. 

We now turn to the non-equilibrium transport. It can be realized by attaching a boundary channel to two electronic reservoirs, 
each being in equilibrium with its own chemical potential and temperature. 
We assume a ballistic boundary channel, which is justified if it is shorter than both elastic and inelastic mean free paths.  
Now, the counter-propagating $\uparrow$ and $\downarrow$ states come from different reservoirs with unequal occupation numbers. 
Say, the $\uparrow$ occupation number is still $f(E)$, while that of the $\downarrow$ mode is 
$f^\prime(E) = \bigr[ {\rm e}^{(E-\mu^\prime)/(k_{_B}T^\prime)} + 1 \bigl]^{-1}$, 
with chemical potential $\mu^\prime \not = \mu$ and temperature $T^\prime \not = T$. 
The net electric current $j = j_\uparrow + j_\downarrow$ can be written as

\begin{eqnarray}
j &=& {\rm sgn}(\epsilon_y) \frac{e}{h}  \int_{E_{\rm v}}^{E_{\rm c}} [f(E) - f^\prime(E)] dE 
\nonumber\\
&\approx& 
{\rm sgn}(\epsilon_y) \biggl[ G \,\, \frac{\mu -\mu^\prime}{e} + G S \,\, (T-T^\prime) \biggr], 
\label{j}
\end{eqnarray}
where we linearized $j$ with respect to the differences $\mu -\mu^\prime$ and $T-T^\prime$ (both assumed small enough), 
introducing the electric conductance, $G$, and the Seebeck coefficient (thermopower), $S$ \cite{Mahan00}:  

\begin{equation}
G = \frac{e^2}{h}   
\int_{E_{\rm v}}^{E_{\rm c}}  \biggl(- \frac{\partial f}{\partial E} \biggr) dE = 
\frac{e^2}{h} [f(E_{\rm v}) - f(E_{\rm c})], 
\label{G}
\end{equation}
\begin{eqnarray}
S &=& \frac{ 1/e }{f(E_{\rm v}) - f(E_{\rm c})}  
\int_{E_{\rm v}}^{E_{\rm c}} \frac{E-\mu}{T} \biggl( -\frac{\partial f}{\partial E} \biggr) dE 
\nonumber\\
&=& \frac{k_{_B}/e}{f(E_{\rm v}) - f(E_{\rm c})}
\int_{ \frac{E_{\rm v} - \mu}{2k_{_B}T} }^{ \frac{E_{\rm c} - \mu}{2k_{_B}T} }  \frac{\eta d\eta}{\cosh^2\eta}.
\label{S}
\end{eqnarray}
Here, the energy bounds $E_{\rm c}$ and $E_{\rm v}$ (\ref{E_cv}) contain 
the details of the band structure of the 1T$^\prime$ - WSe$_2$.
In particular, the transport coefficients reflect the particle-hole asymmetry as well as the atomic and structural SOC.
The overall sign of the current in equation (\ref{j}) depends on the curvature of the particle-hole asymmetric dispersion 
along the $k_y$ direction (see equation (\ref{H_A})). This sign determines which of the two reservoirs acts as the electron source and which as the sink.
The calculation of the transport coefficients is restricted to the subgap states, implying $k_{_B}T < \frac{1}{2} E_g$, 
which holds well up to the room temperatures for 1T$^\prime$ - WSe$_2$ with $E_g \approx 120$ meV.

\begin{figure}[t!]
\begin{center}
\includegraphics[width=75.25mm]{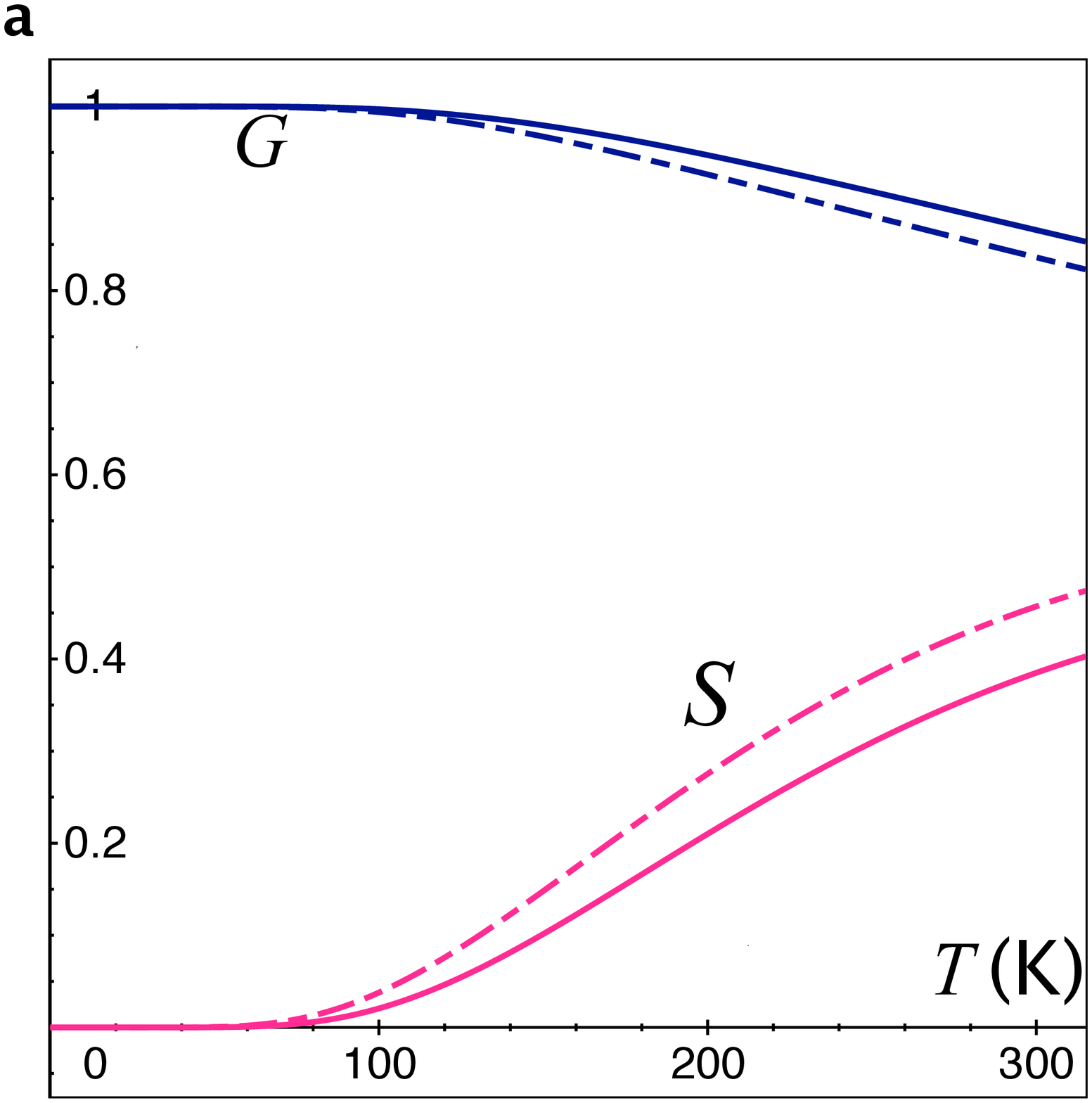}
\includegraphics[width=75.25mm]{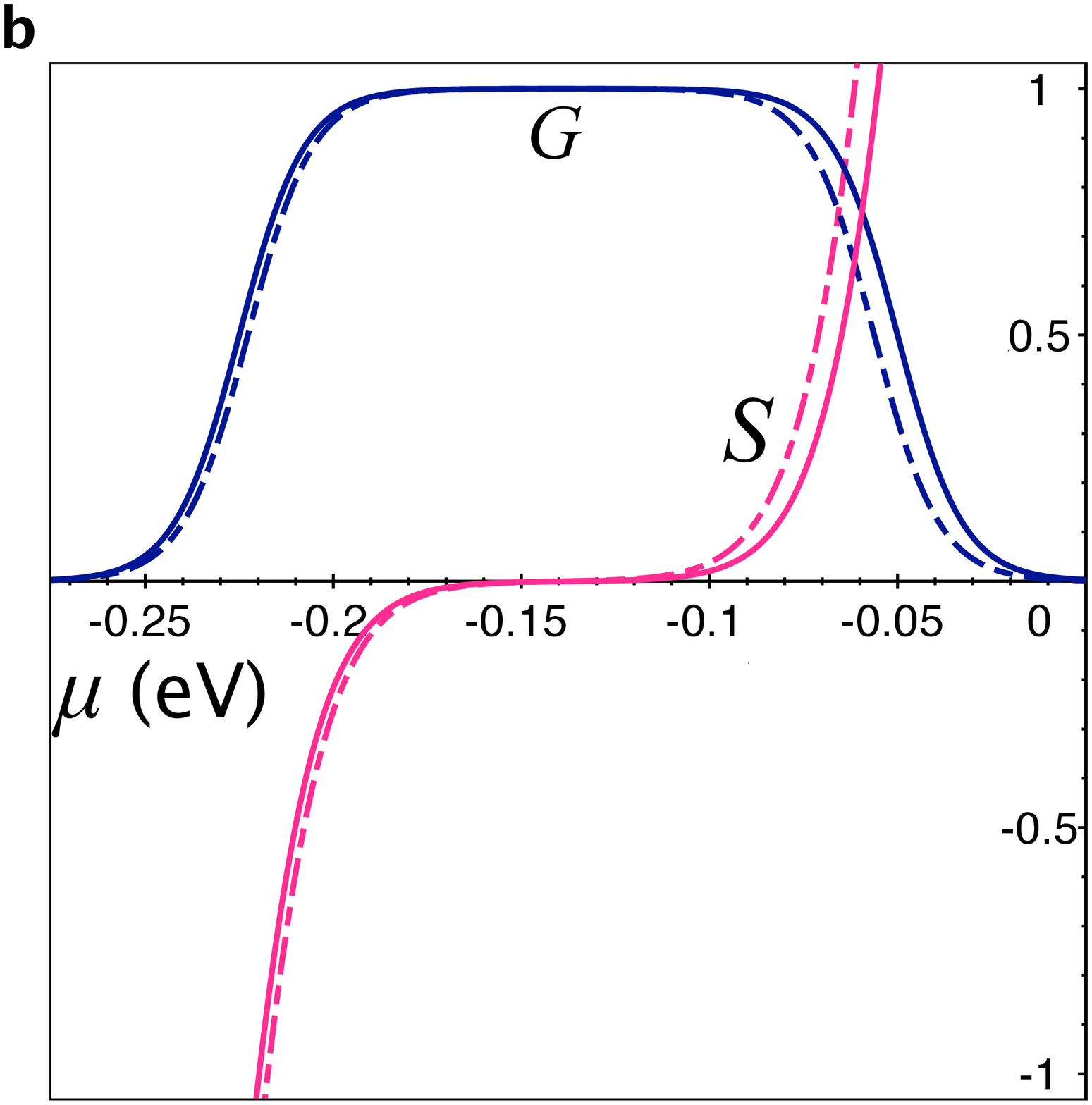}
\end{center}
\caption{
Electric conductance $G$ (in units of $e^2/h$) and thermopower $S$ (in units of $k_{_B}/|e|$) along a 1T$^\prime$ -1H phase boundary 
in WSe$_2$ (see equations (\ref{G}) and (\ref{S})): 
(a) temperature dependence of $G$ and $S$ with the Fermi level in the bandgap, $\mu= -0.1$ eV, and
(b) dependence of thermopower on chemical potential inside the bandgap for $T=100$ K.
Solid and dashed curves are for vanishing and finite structural SOC, with $\lambda=\delta=0$ and 
$\lambda = 0.02$ eV${\rm \AA}$, $\delta = 0.002$ eV, respectively.
Other band structure parameters as the same as in figure \ref{E_WSe2}(a).    
}
\label{G_S}
\end{figure}

Figure \ref{G_S}(a) shows the temperature dependence of equations (\ref{G}) and (\ref{S}). 
The large bandgap of the 1T$^\prime$ - WSe$_2$ manifests itself as the conductance plateau at $e^2/h$ 
up to $\approx 100$ K. The deviation from $e^2/h$ remains less than 20$\%$ up to the room temperatures. 
The thermopower is exponentially suppressed, but grows faster than the conductance deviation from $e^2/h$. 
These observations are corroborated by the asymptotic formulae

\begin{eqnarray}
G &\approx& \frac{e^2}{h} 
\biggl(  
1 - {\rm e}^{ - \frac{\mu - E_{\rm v}}{k_{_B}T} } - {\rm e}^{ - \frac{E_{\rm c} - \mu}{k_{_B}T} }
\biggr), 
\label{G_low}\\
S &\approx& \frac{k_{_B}}{e} 
\biggl( 
\frac{\mu - E_{\rm v}}{k_{_B}T} {\rm e}^{ - \frac{\mu - E_{\rm v}}{k_{_B}T} } - 
\frac{E_{\rm c} - \mu}{k_BT} {\rm e}^{ - \frac{E_{\rm c} - \mu}{k_{_B}T} }
\biggl),
\label{S_low}
\end{eqnarray}
for $k_{_B}T < |E_{\rm c, v} - \mu|$.

As a function of the chemical potential, $G(\mu)$ has a plateau-like maximum close to $e^2/h$ 
\footnote{We note that $G$ does not include the contribution of the bulk states. 
The total conductance of the boundary and bulk states is expected to have a plateau-like minimum at $e^2/h$ \cite{Bernevig06}.
} (see figure \ref{G_S}(b)). 
Away from the center of the plateau  $G(\mu)$ drops exponentially. 
This behaviour is band-structure-dependent, and the knowledge of the spectrum termination points $E_{\rm c}$ and $E_{\rm v}$ 
is the minimal information needed to understand it. 
A specific feature of the thermopower $S(\mu)$ is a sign reversal inside the bandgap of the material. 
The zero of $S(\mu)$ is given by the average of the energy bounds $E_{\rm c}$ and $E_{\rm v}$

\begin{equation}
\mu_0 = \frac{E_{\rm c} + E_{\rm v}}{2} = \epsilon - \frac{\epsilon_y m_0}{m_y},
\label{mu_0}
\end{equation}
which is estimated to be about $-0.14$ eV for 1T$^\prime$ - WSe$_2$. 
The point of the sign reversal $\mu_0$ lies at the center of the conductance plateau and
reflects the particle-hole asymmetry of the band structure. 
Away from the plateau center, the function $S(\mu)$ shows an exponential increase, 
depending on the spectrum termination points $E_{\rm c}$ and $E_{\rm v}$.

Also noteworthy is the effect of the structural inversion asymmetry on the thermoelectric coefficients 
(compare solid and dashed curves in figure \ref{G_S}). It is caused by the shifts of the energy bounds 

\begin{equation}
\Delta E_{\rm c, v} = \delta \, {\rm sgn}(v_x m_x) \mp \lambda k_0, 
\label{dE_cv}
\end{equation}
due to the structural SOC (see equations (\ref{v_epsilon}) and (\ref{E_cv})). 
The effect is well visible already for small values of the SOC constants $\lambda$ and $\delta$ 
such as in typical 2D semiconductor heterostructures. 
This can be explained by the exponential sensitivity of $G$ and $S$ to the changes in the energy bounds  $E_{\rm c}$ and $E_{\rm v}$.
It is worth reminding that we consider the SOC without mixing the spin states.
The Rashba-like SOC \cite{Shi19} requires a separate treatment. 
It can be included in equation (\ref{H_A}), while the particle-hole symmetric part of the Hamiltonian (\ref{H_S}) remains unchanged.
We can therefore still use the approach in \ref{Boundary solution} and expect similar results to those in figure \ref{G_S}.

\section{Discussion and Conclusions}

In fact, the temperature dependence of the boundary conductance has been measured for a related material, 
1T$^\prime$ - WTe$_2$ \cite{Fei17,Wu18}. 
A conductance plateau followed by a decrease in $G(T)$ has been seen in both experiments \cite{Fei17,Wu18}, e.g., in \cite{Wu18} 
the conductance plateau persisted up to 100K. The behaviour of $G(T)$ in figure \ref{G_S}(a) is quite similar,
indicating that the proposed model captures the essential features of the boundary transport. 
Especially, the spectrum termination points of the boundary states have been found important 
for modelling the temperature and chemical potential dependences of the ballistic conductance and thermopower.

Concretely, this work has found that the spectrum termination points determine the boundaries of the conductance plateau 
and the position of the zero of the electric thermopower. We have expressed these features in terms of the bulk band parameters, 
so the behaviour of the boundary transport can be predicted solely on the basis of the bulk band structure. 
These results distinguish the present study from the related previous work 
(see, e.g., Refs. \cite{Takahashi10,Ghaemi10,Takahashi12,Xu14,Xu17,Gusev19}). 

Because of their large bandgap, the 1T$^\prime$ materials should be particularly suitable for measurements of the boundary thermopower, at least conceptually. 
In the model studied above, the thermopower is small for temperatures below the bandgap, but it can be detected by the sign reversal as 
an applied gate voltage shifts the Fermi level between the conduction and valence bands. 

These findings contrast with the prediction of theory \cite{Xu14} invoking energy-dependent scattering times in and outside the bandgap of a 2DTI
(see also review in \cite{Xu17}).    
On the other hand, Gusev {\em et al} \cite{Gusev19} have observed an ambipolar thermopower in a HgTe - based 2DTI system, 
but attributed their findings mainly to the bulk carriers. Regardless of the bulk contribution, 
the transport in 2DTIs typically involves two edges at the opposite sides of the sample. 
A crystalline phase boundary, on the contrary, acts as a single topological channel, 
offering access to still unexplored regimes of topological phases of matter.

\ack
The author thanks Wei Chen for useful discussions.
This work was supported by the German Research Foundation (DFG) through TRR 80. 

\appendix
\section{Solution for topological boundary states}
\label{Boundary solution}

Here we solve the boundary problem posed in Sec. \ref{Boundary states}.
To find the eigenenergy $E$, we integrate equation (\ref{Eq_E}) across the interface,

\begin{equation}
\int^\infty_{-\infty} dx H^{^A}_{\sigma k_y}(x) |x\rangle + \int^\infty_{-\infty} dx H^{^S}_{\sigma k_y}(x) |x\rangle =E \int^\infty_{-\infty} dx |x\rangle,
\label{Eq_E1}
\end{equation}
and evaluate the first integral, using the boundary conditions $\partial_x |x\rangle \to 0$ at $\pm\infty$, which yields  
$(\epsilon_0 + \epsilon_y k^2_y + \lambda \sigma k_y)\int^\infty_{-\infty} dx |x\rangle$.
The second term can be cast into a similar form by choosing $|x\rangle$ to be an eigenstate of the particle-hole symmetric Hamiltonian 
$H^{^S}_{\sigma k_y}(x)$,

\begin{equation}
H^{^S}_{\sigma k_y}(x)|x\rangle = E^{^S}_{\sigma k_y} |x\rangle,
\label{Eq_ES}
\end{equation}
with an eigenvalue $E^{^S}_{\sigma k_y}$. 
Then, collecting the pre-integral factors in equation (\ref{Eq_E1}), we can write the boundary eigenenergy as the sum
\begin{equation}
E_{\sigma k_y} = \epsilon_0 + \epsilon_y k^2_y + \lambda \sigma k_y + E^{^S}_{\sigma k_y}.
\label{E}
\end{equation}
Now, the problem reduces to solving the particle-hole symmetric equation (\ref{Eq_ES}). 
In the 1T$^\prime$ domain ($x \geq 0$), the substitution $|x\rangle = |0\rangle {\rm e}^{-\varkappa x}$ brings equation (\ref{Eq_ES}) to the form 

\begin{equation}
[i\sigma v_x \varkappa \tau_1 - (v_y k_y + \delta\sigma) \tau_2 + (m_0 + m_y k^2_y - m_x\varkappa^2 ) \tau_3]|0\rangle = E^{^S}_{\sigma k_y} |0\rangle.
\label{Eq_ES1}
\end{equation}
Since there are two unknowns, viz. $\varkappa$ and $E^{^S}_{\sigma k_y}$, 
we split (\ref{Eq_ES1}) into two equations

\begin{equation}
[\sigma v_x \varkappa \tau_0 - (m_0 + m_y k^2_y - m_x\varkappa^2 ) \tau_2]|0\rangle = 0, 
\label{Eq_ES2}
\end{equation}
and

\begin{equation}
- (v_y k_y + \delta\sigma) \tau_2 |0\rangle = E^{^S}_{\sigma k_y} |0\rangle, 
\label{Eq_ES3}
\end{equation}
where in the first equation we used $\tau_3 = -i\tau_1\tau_2$. 
Clearly, $|0\rangle$ is an eigenstate of $\tau_2$, so the equations above become algebraic and have the following solutions

\begin{equation}
\varkappa_{1,2} = -\tau \sigma \frac{ v_x }{2m_x } \pm \sqrt{ \left( \frac{ v_x }{2m_x } \right)^2 + \frac{ m_0 + m_y k^2_y}{ m_x } },
\label{kappa_12}
\end{equation}
and 

\begin{equation}
E^{^S}_{\sigma k_y} = - (v_y k_y + \delta\sigma) \tau,
\label{E_S}
\end{equation}
with $\tau$ being an eigenvalue of $\tau_2$. 

Further, the boundary condition $|x\rangle \to 0$ at $+\infty$ only allows the decay constants with a positive real part, ${\rm Re}(\varkappa_{1,2}) > 0$. 
In order to identify those, we recall that the 1T$^\prime$ domain has an inverted band structure (see Table \ref{Table}) 
and, therefore, 

\begin{equation}
\varkappa_1\varkappa_2 = - (m_0 + m_y k^2_y)/m_x \geq 0 
\label{kappa_product}
\end{equation}
for any $k_y$ value in segment (\ref{k_0}).
That is, there is one solution $\varkappa_1$ and one solution $\varkappa_2$, 
each with a positive real part for certain quantum numbers $\tau$ and $\sigma$.
Indeed, if the quantum numbers satisfy the condition 

\begin{equation}
\tau \sigma = -{\rm sgn}(v_x m_x), 
\label{t_s}
\end{equation}
then $\varkappa_1$ and $\varkappa_2$ take the form (\ref{kappa_12_top}) 
with explicit ${\rm Re}(\varkappa_{1,2}) > 0$ for any wave number in segment (\ref{k_0}). 
Accordingly, the boundary solution on the 1T$^\prime$ side has the form

\begin{equation}
|x\rangle = (C_1{\rm e}^{-\varkappa_1 x} + C_2{\rm e}^{-\varkappa_2 x}) |0\rangle,
\label{Sol_1T'}
\end{equation}
where $C_1$ and $C_2$ are constants, while $|0\rangle$ is the eigenstate of $\tau_2$ with eigenvalue $\tau  = -\sigma \, {\rm sgn}(v_x m_x)$ 
(see equation (\ref{t_s})). $\tau$ is chosen such that for a given spin projection $\sigma$ the real parts of $\varkappa_1$ and $\varkappa_2$ are always positive. 
Then, combining equations (\ref{E}), (\ref{E_S}), and (\ref{t_s}), we arrive at the energy dispersion in equation (\ref{Sol_E}) of the main text. 

To complete the calculation, we seek a bound state $|x\rangle = |0\rangle {\rm e}^{\varkappa x}$ in the 1H domain.
The only difference to the 1T$^\prime$ side is the expression for the decay constants  (see also Table \ref{Table}),

\begin{equation}
\varkappa_{3,4} = \tau \sigma \frac{ v_x }{2\overline{m}_x } \pm 
\sqrt{ \left( \frac{ v_x }{2\overline{m}_x } \right)^2 + \frac{ \overline{m}_0 + \overline{m}_y k^2_y}{ \overline{m}_x } }.
\label{kappa_34}
\end{equation}
For the normal 1H band ordering, there is one positive-valued decay constant, $\varkappa_{3}$. 
The quantum numbers  $\tau$ and $\sigma$ are conserved across the interface, which yields the solution in the 1H domain

\begin{equation}
|x\rangle = C_3 {\rm e}^{\varkappa_3 x} |0\rangle.
\label{Sol_1H}
\end{equation}
To obtain equation (\ref{kappa_3}) for $\varkappa_3$ we used $\tau \sigma = -{\rm sgn}(v_x m_x)$ along with
${\rm sgn}(m_x) = - {\rm sgn}(m_0)$ and ${\rm sgn}(\overline{m}_x) = {\rm sgn}(\overline{m}_0)$.
Matching the wave functions (\ref{Sol_1T'}) and (\ref{Sol_1H}) along with their derivatives at $x=0$ leaves one free constant, $C_3$, serving as the normalizing factor:

\begin{equation}
C_1 = - \frac{\varkappa_3 + \varkappa_2}{\varkappa_1 - \varkappa_2} C_3, \qquad 
C_2 =  \frac{\varkappa_3 + \varkappa_1}{\varkappa_1 - \varkappa_2} C_3.
\label{C_12}
\end{equation}
Finally, combining equations (\ref{Sol_1T'}), (\ref{Sol_1H}), and (\ref{C_12}) 
we obtain the bound state in the form (\ref{Sol_x}) discussed in the main text 
(where the constant $C_3$ is absorbed into $|0\rangle$).

\end{document}